\documentclass[journal,draftcls,onecolumn,12pt,twoside]{IEEEtran}
\usepackage{amsmath,amsfonts}
\usepackage{algorithmic}
\usepackage{algorithm}
\usepackage{array}
\usepackage[caption=false,font=footnotesize,labelfont=rm,textfont=rm]{subfig} 
\usepackage{textcomp}
\usepackage{stfloats}
\usepackage{url}
\usepackage{verbatim}
\usepackage{graphicx}
\usepackage{cite}
\usepackage{etoolbox}
\usepackage{xcolor}

\def\BibTeX{{\rm B\kern-.05em{\sc i\kern-.025em b}\kern-.08em
    T\kern-.1667em\lower.7ex\hbox{E}\kern-.125emX}}
\usepackage{balance}

\begin{document}
\title{Anti-jamming Transmission of Downlink Cell Free Millimeter-Wave MIMO System}
\author{Zilong Wang, \textit{Student Member, IEEE}, Cheng Zhang, \textit{Member, IEEE},\\ Changwei Zhang, \textit{Member, IEEE}, Yongming Huang, \textit{Senior Member, IEEE}
\thanks{Z. Wang, C. Zhang and Y. Huang are with the National Mobile Communication Research Laboratory, Southeast University, Nanjing 210096, China, and C. Zhang, C. Zhang and Y. Huang are also with the Purple Mountain Laboratories, Nanjing 211111, China (e-mail: 
zl{\_}wang@seu.edu.cn, zhangcheng{\_}seu@seu.edu.cn, changwei.z@outlook.com, huangym@seu.edu.cn).}
}

\markboth{IEEE Wireless Communications Letters,~Vol.~X, No.~X, June~2024}%
{How to Use the IEEEtran \LaTeX \ Templates}

\maketitle

\begin{abstract}
In this letter, the maximization of resistible jamming power is studied for multi-user downlink millimeter-wave cell-free multiple-input-multiple-output (CF-MIMO) systems. We propose an alternate optimization-based anti-jamming hybrid beamforming (AO-AJHBF) scheme. For receiving beamforming, more practical prior about the jamming channel, i.e., second-order statistics rather than instantaneous information, is exploited via maximizing the generalized Rayleigh quotient. For transmitting beamforming, we use the max-min fairness principle and propose a low-complexity projected gradient ascent-based method to circumvent the excessive computation of semi-definite relaxation (SDR). Simulations verify the performance advantage of proposed AO-AJHBF over schemes based on weighted minimum mean square error and SDR methods.
\end{abstract}
\textbf{Keywords: Cell-free MIMO, millimeter wave, anti-jamming, projected gradient ascent, max-min fairness, hybrid beamforming }

\section{Introduction}  
In recent years, the cell-free (CF) system has gained considerable attention for its advantages in transmission reliability and rate\cite{elhoushy2021cell}. Nevertheless, the issue of anti-jamming remains a significant challenge in CF systems. {There are numerous varieties of jamming, including active jamming and eavesdropping. The active jamming is particularly damaging, as it can impair data transmission and even sever the link. Consequently, it is vital to implement strategies to mitigate jamming effects. The cooperative transmission of distributed access points (APs) of CF systems naturally enhances the anti-jamming performance in the uplink\cite{9686005}. Uplink power control\cite{sabbagh2021cell} and AP selection\cite{nguyen2023pilot} can be effective countermeasures against malicious jamming in CF system. Besides, hybrid beamforming has been proposed as a common feature of millimeter-wave (mmWave) multiple-input-multiple-output (MIMO) systems\cite{8030501}. Some studies achieve jamming rejection through beamforming design in the uplink MIMO system. Unused pilots can be used to estimate the jamming channel and based on which the anti-jamming beamforming is designed\cite{8017521}.}

In the downlink, users are unable to leverage the distributed characteristics for anti-jamming as in the uplink. Consequently, it is imperative to implement effective anti-jamming measures for downlink. Fortunately, the beamforming can effectively be used in this scenario. For the downlink anti-jamming transmission, the best solution is to jointly optimize the receiving beamforming and transmitting beamforming. Considering its complexity, some studies choose to decouple the two, utilizing the null space of the jamming channel to design the receiving beamforming and then using zero-forcing-type transmitting beamforming\cite{zhu2019mitigating},\cite{9310062}. Since it is difficult to obtain perfect jamming channels due to the non-cooperative nature of jammers in practical systems, some studies also utilize the jamming statistical information to design the anti-jamming beamforming, including the classical minimum variance distortionless response (MVDR) method\cite{qi2024anti},\cite{chalise2022optimum}. However, conventional array signal processing methods (e.g., MVDR) focus on beam response rather than rate-dependent metrics, and are not fully applicable to communication anti-jamming where the satisfaction of user rate requirements is particularly important. 

Existing studies on spatial anti-jamming in the downlink mainly focus on the suppression of jamming at the user side. A comprehensive anti-jamming transmission scheme, including the targeted design of both receiving and transmitting beamforming under the constraint of user rate, remains relatively scarce. In addition, the assumption about the jamming channel knowledge is relatively ideal. To address these issues, we investigate a complete anti-jamming transmission scheme for the downlink. The main contributions are as follows:
\begin{itemize}
    \item Contrary to the optimization goals of existing communication anti-jamming efforts (e.g., complete jamming cancellation), we aim at maximizing the resistible jamming power while guaranteeing the user's rate requirements. Meanwhile, more practical channel priori information is considered, i.e., the user side is only informed of the second-order statistics of the jamming channel.
    \item To solve the problem efficiently, we propose an alternate optimization-based anti-jamming hybrid beamforming (AO-AJHBF) framework. Exploiting second-order statistics, the receiving beamforming is designed based on generalized Rayleigh quotient (GRQ). The transmitting beamforming is designed based on the principle of max-min fairness and a low-complexity method based on projected gradient ascent (PGA) is proposed to solve it. 
    \item Simulation results indicate that the proposed scheme outperforms the weighted minimum mean square error (WMMSE) scheme and the semi-definite relaxation (SDR) based scheme in terms of resistible jamming power under different parameters. Furthermore, the impact of antenna number and degree of spatial distribution of jammers on anti-jamming performance is investigated.
\end{itemize}

\section{System Model and Problem Formulation}
Consider a downlink mmWave multi-user CF-MIMO communication system, where $\mathit{K}$ user equipments (UEs) with ${M_{\rm{U}}}$ antennas fully connected with ${M_{\rm{RF}}}$ RF chains are served by $\mathit{L}$ APs connected to a CPU via fronthaul links. Each AP is equipped with $\mathit{M}$ antennas fully connected with ${N_{\rm{RF}}}$ RF chains. Besides, there exist $\mathit{G}$ jammers, each of which is assumed to be equipped with ${M_{\rm{J}}}$ antennas.

\subsection{Channel Model}
Considering the finite scattering properties of the mmWave propagation environment, we utilize the Saleh-Valenzuela channel model, wherein uniform planar arrays (UPAs) are deployed at APs, UEs and jammers, respectively. The array steering vector of an UPA is
\begin{equation} \label{1}
\setlength\abovedisplayskip{1pt}
\begin{array}{c}
{\bf{a}}\left( {\mu ,\nu } \right) = {\left[ {1,{e^{j\frac{{2\pi {d_h}}}{\lambda }\mu }}, \cdots ,{e^{j\frac{{2\pi {d_h}}}{\lambda }\left( {{M_h} - 1} \right)\mu }}} \right]^{\rm{T}}} \otimes \\
{\left[ {1,{e^{j\frac{{2\pi {d_v}}}{\lambda }\nu }}, \cdots ,{e^{j\frac{{2\pi {d_v}}}{\lambda }\left( {{M_v} - 1} \right)\nu }}} \right]^{\rm{T}}},
\end{array}
\setlength\belowdisplayskip{1pt}
\end{equation}
where ${M_h}$ and ${M_v}$ are the number of horizontal and vertical antennas. ${d_h}$ and ${d_v}$ are the horizontal and vertical antenna
distances. $\mu  = \sin \theta \sin \varphi $ and $\nu  = \cos \theta $ denote virtual angles. $\varphi $, $\theta $ and $\lambda $ are the azimuth angle, the elevation angle and the wavelength, respectively.
 
Assuming that the number of downlink multipath components between AP $\mathit{l}$ and UE $\mathit{k}$ is ${{P_{l,k}}}$. The channel between AP $\mathit{l}$ and UE $\mathit{k}$ ${{\bf{H}}_{l,k}}\in {\mathbb{C}^{{M_{\rm{U}}} \times M}}$ is expressed as
\begin{equation}\label{2}
\setlength\abovedisplayskip{1pt}
{{\bf{H}}_{l,k}} = \sum\nolimits_{p = 1}^{{P_{l,k}}} {{\alpha _{l,k}^{p}}} {\sqrt{\beta _{l,k}^{p}}}{{\bf{a}}_{\rm{U}}}({{\mathord{ \mu } }_{l,k}^{p}},{{\mathord{ \nu } }_{l,k}^{p}}){\bf{a}}_{\rm{A}}^{\rm{H}}({{\mathord{ \mu } }_{k,l}^{p}},{{\mathord{ \nu } }_{k,l}^{p}}),
\setlength\belowdisplayskip{1pt}
\end{equation}
where ${\alpha _{l,k}^{p}}$ is small scale fading and ${{\beta _{l,k}^{p}}}$ is large scale fading of ${p}$th path. And the channel with ${{{\bar P}_{g,k}}}$ paths between jammer $\mathit{g}$ and UE $\mathit{k}$ ${{\bf{J}}_{g,k}}\in {\mathbb{C}^{{M_{\rm{U}}} \times {M_{\rm J}}}}$ is expressed as
\begin{equation}\label{3}
\setlength\abovedisplayskip{1pt}
{{\bf{J}}_{g,k}} = \sum\nolimits_{p = 1}^{{{\bar P}_{g,k}}} {{{\bar \alpha }_{g,k}^{p}}}{\sqrt{{\bar \beta }_{g,k}^{p}}} {{\bf{a}}_{\rm{U}}}({{\mathord{\mu } }_{g,k}^{p}},{{\mathord{ \nu } }_{g,k}^{p}}){\bf{a}}_{\rm{J}}^{\rm{H}}({{\mathord{ \mu } }_{k,g}^{p}},{{\mathord{ \nu } }_{k,g}^{p}}),
\setlength\belowdisplayskip{1pt}
\end{equation}
where ${\bar \alpha _{g,k}^{p}}$ is small scale fading and ${{{\bar \beta }_{g,k}^{p}}}$ is large scale fading of ${p}$th path. ${{\bf{a}}_{\rm{U}}}$, ${{\bf{a}}_{\rm{A}}}$ and ${{\bf{a}}_{\rm{J}}}$ are the array steering vectors of UE, AP and jammer, respectively.

\newcounter{TempEqCnt} 
\setcounter{TempEqCnt}{\value{equation}} 
\setcounter{equation}{3} 
\begin{figure*}[b] 
\centering 
\hrulefill 
\begin{equation}\label{4}
\begin{array}{c}
{y_k} = \underbrace {\sum\nolimits_{l = 1}^L {{\bf{w}}_k^{\rm{BB,H}}{\bf{W}}_k^{\rm{RF,H}}{{{\bf{\bar H}}}_{l,k}}{\bf{F}}_l^{\rm{RF}}{\bf{f}}_{l,k}^{\rm{BB}}{x_k}} }_{{\rm{Known~ Useful~Signal}}} + \underbrace {\sum\nolimits_{j \ne k}^K {\sum\nolimits_{l = 1}^L {{\bf{w}}_k^{\rm{BB,H}}{\bf{W}}_k^{\rm{RF,H}}{{{\bf{\bar H}}}_{l,k}}{\bf{F}}_l^{\rm{RF}}{\bf{f}}_{l,j}^{\rm{BB}}{x_j}} } }_{{\rm{Known~Multi - user~Interference}}} + \\ \underbrace {\sum\nolimits_{j = 1}^K {\sum\nolimits_{l = 1}^L {{\bf{w}}_k^{\rm{BB,H}}{\bf{W}}_k^{\rm{RF,H}} ({{{\bf{\tilde H}}}_{l,k}}+{{{\bf{\Sigma }}_{l,k}}})
{\bf{F}}_l^{\rm{RF}}{\bf{f}}_{l,j}^{\rm{BB}}{x_j}} } }_{{\rm{Channel~Estimate~and~Quantization~Error}}}
 + \underbrace {\sum\nolimits_{g = 1}^G {\sqrt {{q_{g,k}}} {\bf{w}}_k^{\rm{BB,H}}{\bf{W}}_k^{\rm{RF,H}}{\bf{J}}_{g,k}{{\bf{w}}_{{\rm J},g,k}}{z_g}} }_{{\rm{Jamming}}}\\ 
 + \underbrace {{\bf{w}}_k^{\rm{BB,H}}{\bf{W}}_k^{\rm{RF,H}}{{\bf{n}}_k}}_{{\rm{Receiver~Noise}}}
\end{array}
\end{equation}
\end{figure*}
\setcounter{equation}{\value{TempEqCnt}} 

\subsection{Transmission Model}
First, to facilitate processing, denote the entire channel matrix ${{\bf{H}}_{k}} = \left[ {{\bf{H}}_{1,k},{\bf{H}}_{2,k}, \cdots ,{\bf{H}}_{L,k}} \right] \in {\mathbb{C}^{{M_{\rm{U}}} \times \left( {L \times M} \right)}}$ for UE ${k}$. Also denote the entire analog beamforming matrix ${{\bf{F}}^{{\rm{RF}}}} = blkdiag$ $\left( {{\bf{F}}_1^{{\rm{RF}}},{\bf{F}}_2^{{\rm{RF}}}, \cdots ,{\bf{F}}_L^{{\rm{RF}}}} \right) \in {\mathbb{C}^{\left( {L \times M} \right) \times \left( {L \times {N_{\rm{RF}}}} \right)}}$ and entire digital beamforming vector ${\bf{f}}_{k}^{{\rm{BB}}} = [ {\bf{f}}_{1,k}^{{\rm{BB,H}}}, $ ${\bf{f}}_{2,k}^{{\rm{BB,H}}}, \cdots ,{\bf{f}}_{L,k}^{{\rm{BB,H}}} ]^{\rm{H}} \in {\mathbb{C}^{\left({L \times {N_{\rm{RF}}}}\right) \times 1}}$, where ${\bf{F}}_l^{{\rm{RF}}} \in {\mathbb{C}^{M \times {N_{\rm{RF}}}}}$ is the analog beamforming at ${l}$th AP and ${\bf{f}}_{l,k}^{{\rm{BB,H}}} \in {\mathbb{C}^{{N_{\rm{RF}}}\times 1}}$ is the digital beamforming from ${l}$th AP to ${k}$th UE. The data for UE ${k}$ denoted by ${x_k}$ satisfies ${\rm E}\left\{ {{x_k}x_k^*} \right\} = 1$ and ${\rm E}\left\{ {{x_k}x_j^*} \right\} = 0,\forall j \ne k$. For the hybrid beamforming scheme, the data signal is firstly precoded by the digital beamforming vector ${\bf{f}}_{k}^{{\rm{BB}}}$, and then processed by the analog beamforming ${{\bf{F}}^{{\rm{RF}}}}$. Assuming that the signals are perfectly synchronised at the UE, then received signal is first processed by analog combiner ${\bf{W}}_k^{\rm{RF,H}}\in {\mathbb{C}^{{M_{\rm{RF}}} \times {M_{\rm{U}}}}}$ at UE ${k}$. Finally, the original signal can be decoded by digital combiner ${{\bf{w}}_{k}^{{\rm{BB,H}}}}\in {\mathbb{C}^{1 \times {M_{\rm{RF}}}}}$. Besides, ${q_{g,k}}$ and ${{\bf{w}}_{{\rm{J}},g,k}}$ are the jamming power and the full-digital beamforming from ${g}$th jammer to ${k}$th UE. ${z_{g}}$ is the jammer signal that satisfies ${\rm E}\left\{ {{z_{g}}z_{g}^*} \right\} = 1$. Additionally, ${{\bf{n}}_k}\sim{\mathcal{CN}}\left( {0,{\sigma ^2}{\bf{I}}} \right)$ is an additive white Gaussian noise vector for UE ${k}$. Then, the received signal of UE ${k}$ can be expressed as \eqref{4}. {The estimated channel is expressed as ${{\bf{\hat H}}_{l,k}}$ and the estimation error is presented as ${{\bf{\tilde H}}_{l,k}}$. By using MMSE estimation, we can get the estimated channel as ${{\bf{\hat h}}_{l,k}} = vec\left( {{{{\bf{\hat H}}}_{l,k}}} \right) = {{\bf{R}}_{l,k}}{\bf{\tilde F}}_k^{\rm{H}}{\bf{\Psi }}_{l,k}^{ - 1}{{\bf{y}}_{l,k}}$, where ${\bf{\tilde F}}_k^{} = {\bf{F}}_k^{\rm{T}} \otimes {\bf{I}}$, ${\bf{\Psi }}_{l,k}^{} = {\tau _p}{\bf{\tilde F}}_k^{}{{\bf{R}}_{l,k}}{\bf{\tilde F}}_k^{\rm{H}} + {\sigma ^2}{\bf{I}}$, ${{\bf{y}}_{l,k}} = vec\left( {{{\bf{Y}}_{l,k}}} \right) = {\tau _p}{\bf{\tilde F}}_k^{}{{\bf{h}}_{l,k}} + {{\bf{n}}_l}$ and ${{\bf{n}}_l} = vec\left( {{{\bf{N}}_l}{\bf{\Phi }}_k^*} \right)$. ${{\bf{N}}_l} \sim CN\left( {0,{\sigma ^2}{\bf{I}}} \right)$. ${{\bf{F}}_k}$ is the precoding of UE-$k$ when sending the pilot. And the pilot matrix $\Phi _k$ satisfies $\Phi _k^{\rm{H}}{\Phi _k} = {\tau _p}{\bf{I}}$ and $\Phi _k^{\rm{H}}{\Phi _j} = {\bf{0}}, j \ne k$. The estimation error ${{\bf{\tilde h}}_{l,k}} = {{\bf{h}}_{l,k}} - {{\bf{\hat h}}_{l,k}}$ and ${{\bf{\hat h}}_{l,k}}$ are independent. And we can get the auto-relation matrix of ${{\bf{\hat h}}_{l,k}}$, which is ${{\bf{\hat R}}_{l,k}} \buildrel \Delta \over = {\tau _p}{{\bf{R}}_{l,k}}{\bf{\tilde F}}_k^{\rm{H}}{\bf{\Psi }}_{l,k}^{ - 1}{\bf{\tilde F}}_k^{}{{\bf{R}}_{l,k}}$. Then, the auto-relation matrix of ${{\bf{\hat h}}_k}$ is {{${{\bf{\hat R}}_k} = blkdiag\left( {{{{\bf{\hat R}}}_{1,k}},{{{\bf{\hat R}}}_{2,k}}, \cdots ,{{{\bf{\hat R}}}_{L,k}}} \right)$}}. Finally, the auto-relation matrix of estimation error ${{\bf{\tilde h}}_k}$ is ${{\bf{Q}}_k} = {{\bf{R}}_k} - {{\bf{\hat R}}_k}$. Furthermore, considering channel estimation errors and quantization errors caused by limited capacity of fronthaul, quantized channel is expressed as ${{\bf{\bar H}}_{l,k}={\mathrm Q}({{\bf{\hat H}}_{l,k}})=\alpha{{\bf{\hat H}}_{l,k}}+{\bf{\Sigma }}_{l,k}}$ and the quantization error is ${{\bf{\Sigma }}_{l,k}}$. ${vec({\bf{\Sigma }}_{l,k})}\sim{\mathcal{CN}}\left( {0,{\sigma_{q,l,k} ^2}{\bf{I}}} \right)$.  {{$\sigma _{q,l,k}^2 = \alpha \left( {1 - \alpha } \right){\tau _p}{\rho _p}\beta _{l,k}^2{\left( {{\tau _p}{\rho _p}\sum\nolimits_{j \ne k}^K {{\beta _{l,j}}{{\left| {\Phi _k^{\rm{H}}{\Phi _j}} \right|}^2}}  + 1} \right)^{ - 1}}$}} is relevant to distortion factor $\alpha$ and large scale coefficient (more details about quantization can be referred to Appendix A). Denote desired signal power, MU interference power, interference power caused by channel estimation error, quantization error, jamming power and noise power as ${{\rm{DS}}_k}$, ${\rm{I}}{{\rm{S}}_k}$, ${\rm{E}}{{\rm{N}}_k}$, ${\rm{Q}}{{\rm{E}}_k}$, ${\rm{J}}{{\rm{S}}_k}$, ${\rm{N}}{{\rm{S}}_k}$, respectively, i.e.,} 
\setcounter{equation}{4} 
\begin{subequations}\label{5}
\setlength\abovedisplayskip{1pt}
\begin{align}
{\rm{D}}{{\rm{S}}_k} &= {\left| {{\bf{w}}_{k}^{{\rm{BB}},{\rm{H}}}{\bf{W}}_k^{{\rm{RF}},{\rm{H}}}{{{\bf{\bar H}}}_k}{{\bf{F}}^{{\rm{RF}}}}{\bf{f}}_k^{{\rm{BB}}}} \right|^2}\\
{\rm{I}}{{\rm{S}}_k} &= \sum\nolimits_{j \ne k}^K {{{\left| {{\bf{w}}_k^{{\rm{BB}},{\rm{H}}}{\bf{W}}_k^{{\rm{RF}},{\rm{H}}}{{{\bf{\bar H}}}_k}{{\bf{F}}^{{\rm{RF}}}}{\bf{f}}_j^{{\rm{BB}}}} \right|}^2}}\\
{\rm{E}}{{\rm{N}}_k} &= \sum\nolimits_{j = 1}^K {{\rm{E}}\left\{ {{{\left| {{\bf{w}}_k^{{\rm{BB}},{\rm{H}}}{\bf{W}}_k^{{\rm{RF}},{\rm{H}}}{{{\bf{\tilde H}}}_k}{{\bf{F}}^{{\rm{RF}}}}{\bf{f}}_j^{{\rm{BB}}}} \right|}^2}} \right\}} \\
{\rm{Q}}{{\rm{E}}_k} &= \sum\nolimits_{j = 1}^K {{\rm{E}}\left\{ {{{\left| {{\bf{w}}_k^{{\rm{BB}},{\rm{H}}}{\bf{W}}_k^{{\rm{RF}},{\rm{H}}}{{{\bf{\Sigma }}_{k}}}{{\bf{F}}^{{\rm{RF}}}}{\bf{f}}_j^{{\rm{BB}}}} \right|}^2}} \right\}} \\
{\rm{J}}{{\rm{S}}_k} &= \sum\nolimits_{g = 1}^G {{q_{g,k}}{\bf{w}}_k^{{\rm{BB}},{\rm{H}}}{\bf{W}}_k^{{\rm{RF}},{\rm{H}}}{{\bf{R}}_{g,k}}} {\bf{W}}_k^{{\rm{RF}}}{\bf{w}}_k^{{\rm{BB}}}\\
{\rm{N}}{{\rm{S}}_k} &= \sigma _k^2\left\| {{\bf{w}}_k^{{\rm{BB}},{\rm{H}}}{\bf{W}}_k^{{\rm{RF}},{\rm{H}}}} \right\|_2^2.
\end{align}
\setlength\belowdisplayskip{1pt}
\end{subequations}

The beamforming design requires only the second-order statistics of the jamming channel, which is given by ${{\bf{R}}_{g,k}} = {\rm{E}}\left\{ {{{\bf{J}}_{g,k}}{{\bf{w}}_{{\rm J},g,k}}{\bf{w}}_{{\rm J},g,k}^{\rm{H}}{{{{{\bf{J}}_{g,k}^{\rm{H}}}} }}} \right\}$. Then the signal-to-interference-plus-noise ratio (SINR) of UE ${k}$ can be expressed as
\begin{equation}\label{6}
\setlength\abovedisplayskip{1pt}
{{\mathop{\rm SINR}\nolimits} _{k}} = \frac{{{{{\mathop{\rm DS}\nolimits} }_{k}}}}{{{{{\mathop{\rm IS}\nolimits} }_{k}} + {\rm{E}}{{\rm{N}}_{k}} +{\rm{Q}}{{\rm{E}}_{k}}+ {{{\mathop{\rm JS}\nolimits} }_{k}}+{{{\mathop{\rm NS}\nolimits} }_{k}}}}.
\setlength\belowdisplayskip{1pt}
\end{equation}

\subsection{Jamming Strategy}
To explore spatial anti-jamming capacity, we consider full-band repressive jammers that disrupt a system in all frequency bands of signal transmission. And each jammer is assumed to possess full-digital beamforming capability and perfect channel state information (CSI) between each UE and itself, which represents the worst-case scenario for the system. It is assumed that the MRT scheme is adopted by jammers. Using singular value decomposition (SVD), the channel matrix between jammer ${g}$ and UE ${k}$ can be expressed as
\begin{equation}\label{7}
\setlength\abovedisplayskip{1pt}
{\bf{J}}_{g,k} = {\bf{U}}_{g,k}\Sigma _{g,k}{{{\bf{V}}_{g,k}^{\rm{H}}}},
\setlength\belowdisplayskip{1pt}
\end{equation}
where ${\bf{V}}_{g,k} = \left[ {{\bf{v}}_{g,k,1},{\bf{v}}_{g,k,2}, \cdots ,{\bf{v}}_{g,k,{M_{\rm J}}}} \right]$. Then, the normalized ${{\bf{v}}_{g,k,1}}$ is chosen as the full-digital beamforming ${{\bf{w}}_{{\rm{J}},g,k}}$.

\subsection{Problem Formulation}
It is of paramount importance to consider rate guarantees in communication anti-jamming. Therefore, we model the problem as a form of maximizing the resistible jamming power while guaranteeing the SINR of each UE. The corresponding formulation is written as

\begin{subequations}\label{8}
\setlength\abovedisplayskip{1pt}
\begin{align}  
\label{8a} &\mathop {\max }\limits_{\scriptstyle\left\{ {{\bf{w}}_k^{{\rm{BB}}}} \right\},\left\{ {{\bf{W}}_k^{\rm{RF}}} \right\},\left\{ {{\bf{F}}_l^{\rm{RF}}} \right\},\hfill\atop
\scriptstyle \left\{ {{\bf{f}}_{l,k}^{{\rm{BB}}}} \right\},\left\{ {{q_{g,k}}} \right\}\hfill} \sum\nolimits_{g = 1}^G {\sum\nolimits_{k = 1}^K {{q_{g,k}}} } \\ 
\label{8b} \rm{s.t.}&\:\: {{\mathop{\rm SINR}\nolimits} _{k}} \ge {\gamma _{th }},\forall k \in K, \\ 
\label{8c} &\left\| {{\bf{w}}_{k}^{{\rm{BB,H}}}{\bf{W}}_k^{{\rm{RF,H}}}} \right\|_2^2 = 1,\forall k \in K,\\ 
\label{8d}  &\sum\nolimits_{k = 1}^K \left\| {{\bf{F}}_l^{{\rm{RF}}}{\bf{f}}_{l,k}^{\rm BB}} \right\|_2^2 \le {{P_{\max }}},\forall l \in L, \\ 
\label{8e}  &\left| {{\bf{F}}_l^{{\rm{RF}}}\left( {m,n} \right)} \right| = 1,\left| {{\bf{W}}_k^{{\rm{RF}}}\left( {m,n} \right)} \right| = 1,\forall m,n,
\end{align}
\end{subequations}
where \eqref{8b} is given to ensure the threshold of SINR of UE. \eqref{8c} represents the normalization of the received combining, as it has no impact on the SINR. \eqref{8d} is the constraints of the power per AP and \eqref{8e} states the constant amplitude constraint of analog beamforming. Obviously, \eqref{8} is a complex non-linear programming problem, which is NP-hard. To solve this problem, we divide the problem into beamforming design and jamming power solving based on AO. Besides, to simplify the problem, the optimal full-digital beamforming vectors (${{\bf{w}}_{k}^{{\rm{FD}}}}={{\bf{W}}_{k}^{{\rm{RF}}}}{{\bf{w}}_{k}^{{\rm{BB}}}}$ and ${\bf{f}}_{k}^{{\rm{FD}}} = {{\bf{F}}^{{\rm{RF}}}}{\bf{f}}_{k}^{{\rm{BB}}}$) are firstly calculated and subsequently transformed into hybrid beamforming.

\section{Alternate Optimization Based Anti-jamming Hybrid Beamforming Design}
In this section, an AO method is proposed to solve the problem. The problem is decomposed into receiving beamforming design, transmitting beamforming design and maximization resistible jamming power, which can be solved alternately.

\subsection{Receiving Beamforming}\label{AA}
For this sub-problem, given the transmitting beamforming and jamming power, the design of each UE's receiving beamforming is independent from the others. Obviously, a higher SINR for the UE is more resilient to jamming. Therefore, it can be represented as ${K}$ separate sub-problems as
\begin{equation}\label{9}
\setlength\abovedisplayskip{1pt}
    \begin{aligned}
      \mathop {\max }\limits_{\left\{ {{\bf{w}}_{k}^{{\rm{FD}}}} \right\}} {{\mathop{\rm SINR}\nolimits} _{k}},~~ 
      \rm{s.t.} \:\:& \left\| {{\bf{w}}_{k}^{{\rm{FD}}}} \right\|_2^2 = 1.
    \end{aligned}
\setlength\belowdisplayskip{1pt}
\end{equation}

{To facilitate the calculation, we choose to optimize the lower bound of SINR. ${{\rm EN}_{k}}$ and ${{\rm QE}_{k}}$ in \eqref{6} can be reformulated as
\begin{equation}\label{10}
\setlength\abovedisplayskip{1pt}
\begin{array}{l}
{\rm{E}}{{\rm{N}}_{k}} \le  LK{P_{\max }} \cdot {\lambda _{\max }}\left( {{{\bf{Q}}_k}} \right) = {\rm{E}}{{\rm{N}}^{\rm UB}_{k}}\\
{\rm{Q}}{{\rm{E}}_{k}} \le  LK{P_{\max }} {\omega _k} = {\rm{QE}}{^{\rm UB}_{k}},
\end{array}
\setlength\belowdisplayskip{1pt}
\end{equation}
where ${{\bf{Q}}_{k}} = {\mathop{\rm E}\nolimits} \left\{ {vec\left( {{{{\bf{\tilde H}}}_{k}}} \right)ve{c^{\rm{H}}}\left( {{{{\bf{\tilde H}}}_{k}}} \right)} \right\}$. More details about the proof of \eqref{10} can be referred to Appendix A}. Then the ${{\mathop{\rm SINR}\nolimits} _{k}^{\rm LB}}$ of UE ${k}$ can be expressed as
\setlength\abovedisplayskip{1pt}
\begin{equation}\label{11}
    {{\mathop{\rm SINR}\nolimits} _{k}} \ge \frac{{{\bf{w}}_k^{\rm FD,H}{{\bf{A}}_k}{\bf{w}}_k^{\rm FD}}}{{{\bf{w}}_k^{\rm FD,H}{{\bf{B}}_k}{\bf{w}}_k^{\rm FD}}} = {{\mathop{\rm SINR}\nolimits} _{k}^{\rm LB}} = {\xi_{k}},
\setlength\belowdisplayskip{1pt}
\end{equation}
where ${{\bf{A}}_k} \succ 0$, ${{\bf{B}}_k} \succ 0$ and 
\begin{equation}\label{12}
\setlength\abovedisplayskip{1pt}
\begin{aligned}
    {{\bf{A}}_k} &= {{\bf{\bar H}}_{k}}{\bf{f}}_{k}^{\rm FD}{\bf{f}}_{k}^{\rm FD,H}{\bf{\bar H}}_{k}^{\rm H}\\
    {{\bf{B}}_k} &= \sum\nolimits_{j \ne k}^K {{{\bf{\bar H}}_{k}}{\bf{f}}_{j}^{\rm FD}{\bf{f}}_{j}^{\rm FD,H}{\bf{\bar H}}_{k}^{\rm H}}    + \sum\nolimits_{g = 1}^G {{q_{g,k}}{{\bf{R}}_{g,k}}}\\
    &+ LK{P_{\max }} \cdot [{\lambda _{\max }}\left( {{{\bf{Q}}_k}} \right)+{\omega _k}]{{\bf{I}}}+ \sigma _{k}^2{{\bf{I}}}.
\end{aligned}
\setlength\belowdisplayskip{1pt}
\end{equation} 

Exploiting the solution of GRQ, it is convenient to derive the receiving beamforming, which is the eigenvector corresponding to the maximum eigenvalue of ${{\bf{B}}_k^{ - 1}{{\bf{A}}_k}}$, i.e.,
\setlength\abovedisplayskip{1pt}
\begin{equation}\label{13}
    {\bf{w}}_{k}^{{\rm{FD}}} = {v_{\max }}\left( {{\bf{B}}_k^{ - 1}{{\bf{A}}_k}} \right).
\setlength\belowdisplayskip{1pt}
\end{equation}
The complexity of multiplication and eigenvalue computation in \eqref{12} is $O(L^2M^2M_{\rm U})$, and the complexity of inverse operation, multiplication and eigenvalue computation in \eqref{13} is $O(M^3_{\rm U}+L^2M^2M_{\rm U})$. The complexity of this method is $O(M^3_{\rm U}+L^2M^2M_{\rm U})$ for each UE.

\subsection{Transmitting Beamforming}
After determining the receiving beamforming, we proceed to solve for the transmitting beamforming. To guarantee data rate of each UE, we focus on the max-min fairness problem at the AP side, which can be defined as
\begin{subequations}\label{14}
\setlength\abovedisplayskip{1pt}
\begin{align}
\label{14a}&\mathop {\max }\limits_{\left\{ {{\bf{f}}_{k}^{\rm{FD}}} \right\}} \mathop {\min }\limits_k {{\xi_{k}}} \\
\label{14b}\rm{s.t.} \:\: &\sum\nolimits_{k = 1}^K {\left\| {{\bf{f}}_{l,k}^{\rm FD}} \right\|_2^2}  \le {P_{\max }},\forall l \in L.
\end{align}
\end{subequations}

Using the SDR method\cite{luo2010semidefinite}, we can transform this non-convex problem into a convex one, which is represented as
\begin{equation}\label{15}
\setlength\abovedisplayskip{1pt}
\begin{aligned}
&\mathop {\max }\limits_{\left\{ {{\bf{U}}_{k}^{\rm{FD}}} \right\}, {{\varepsilon }} } {\varepsilon } \\
{\rm s.t.}\:\:&\sum\nolimits_{k = 1}^K {Tr\left\{ {{\bf{U}}_k^{{\rm{FD}}}{{\bf{S}}_l}} \right\}}  \le {P_{\max }},\forall l \in L,\\
&{\bf{U}}_{k}^{{\rm{FD}}} \succeq 0,{\rm{rank}}\left( {{\bf{U}}_{k}^{\rm{FD}}} \right) = 1,{\xi_{k}} \ge \varepsilon , \forall k \in K,
\end{aligned}
\setlength\belowdisplayskip{1pt}
\end{equation}
where ${\bf{U}}_{k}^{{\rm{FD}}} = {\bf{f}}_{k}^{\rm{FD}}{\bf{f}}_{k}^{\rm{FD,H}}$. And ${{{\bf{S}}_l}}$ is a selective matrix, of which only the $(l-1)M+1$ through $lM$ elements of the main diagonal are 1, the rest are 0. Except for the restriction that ${\rm{rank}}\left( {{\bf{U}}_{k}^{\rm{FD}}} \right) = 1$, the whole problem is convex. 

Interior point method is a classical method to solve \eqref{15}, and the complexity is $O(K^{3}L^{6.5}M^{6.5}{\rm log({1/\epsilon})})$ (where ${\epsilon}$ is the accuracy of the solution) \cite{ye2011interior}, which is extremely high when the number of antennas is large. Besides, the optimal solution of \eqref{14} is hardly obtained by SDR based method. Therefore, it is necessary for a low-complexity way to solve this problem. Since the function $min_k$ in \eqref{14} is not differentiable, it is hard to solve the problem directly. Fortunately, a softmax operator can be applied, which can transfer the function $min_k$ into a differentiable one\cite{asadi2017alternative}. The problem \eqref{14a} is reformulated as
\begin{equation}\label{16}
\begin{array}{c}
\mathop {\max }\limits_{\left\{ {{\bf{f}}_k^{{\rm{FD}}}} \right\}} \eta  \left( {{\bf{f}}_k^{{\rm{FD}}}} \right) =  \frac{{\sum\nolimits_{k = 1}^K {{\xi_{k}}{e^{\delta {\xi_{k}}}}} }}{{\sum\nolimits_{k = 1}^K {{e^{\delta {\xi_{k}}}}} }},~~{\rm s.t.} ~~\eqref{14b}
\end{array}
\end{equation}
where the function $\eta$ is concave and the constraint of power is convex. ${\delta \rightarrow - \infty}$ is a parameter, and a negative number less than -3 is sufficient in practice. Naturally, the PGA method can be exploited to solve this problem.

The gradient of $\eta$ over ${{\bf{f}}_k^{{\rm{FD}}}}$ can be expressed as
\begin{equation}\label{17}
    {\nabla _{{\bf{f}}_k^{{\rm{FD}}}}}\eta = \sum\nolimits_{\tilde k = 1}^K {{\nabla _{{\xi _{\tilde k}}}}\eta \cdot {\nabla _{{\bf{f}}_k^{{\rm{FD}}}}}{\xi _{\tilde k}}} ,
\end{equation}
where
\begin{equation}\label{18}
\setlength\abovedisplayskip{1pt}
\begin{aligned}
        &{\nabla _{{\xi _{\tilde k}}}} \eta = \\
        &\frac{{\left( {1 + \delta {\xi _{\tilde k}}} \right){e^{\delta {\xi _{\tilde k}}}}\left( {\sum\nolimits_{\tilde k = 1}^K {{e^{\delta {\xi _{\tilde k}}}}} } \right) - \delta {e^{\delta {\xi _{\tilde k}}}}\left( {\sum\nolimits_{\tilde k = 1}^K {{\xi _{\tilde k}}{e^{\delta {\xi _{\tilde k}}}}} } \right)}}{{{{\left( {\sum\nolimits_{\tilde k = 1}^K {{e^{\delta {\xi _{\tilde k}}}}} } \right)}^2}}},
\end{aligned}
\setlength\belowdisplayskip{1pt}
\end{equation}
and
\begin{equation}\label{19}
\setlength\abovedisplayskip{1pt}
\begin{aligned}
    &{\nabla _{{\bf{f}}_k^{{\rm{FD}}}}}{\xi _{\tilde k}} = \\
    &\left\{ 
{\begin{array}{*{20}{c}}
{\frac{{{\bf{\bar H}}_{\tilde k}^{\rm{H}}{\bf{w}}_{\tilde k}^{{\rm{FD}}}{\bf{w}}_{\tilde k}^{{\rm{FD,H}}}{{{\bf{\bar H}}}_{\tilde k}}{\bf{f}}_k^{{\rm{FD}}}}}{{\sum\nolimits_{j \ne \tilde k}^K {{{\left| {{\bf{w}}_{\tilde k}^{{\rm{FD,H}}}{{{\bf{\bar H}}}_{\tilde k}}{\bf{f}}_j^{{\rm{FD}}}} \right|}^2}}  + \zeta}},\tilde k = k}\\
{\frac{{ - {{\left| {{\bf{w}}_{\tilde k}^{{\rm{FD,H}}}{{{\bf{\bar H}}}_{\tilde k}}{\bf{f}}_{\tilde k}^{{\rm{FD}}}} \right|}^2}{\bf{\bar H}}_{\tilde k}^{\rm{H}}{\bf{w}}_{\tilde k}^{{\rm{FD}}}{\bf{w}}_{\tilde k}^{{\rm{FD,H}}}{{{\bf{\bar H}}}_{\tilde k}}{\bf{f}}_k^{{\rm{FD}}}}}{{{{\left( {\sum\nolimits_{j \ne \tilde k}^K {{{\left| {{\bf{w}}_{\tilde k}^{{\rm{FD,H}}}{{{\bf{\bar H}}}_{\tilde k}}{\bf{f}}_j^{{\rm{FD}}}} \right|}^2}}  + \zeta} \right)}^2}}},\tilde k \ne k},
\end{array}}
\right.
\end{aligned}
\setlength\belowdisplayskip{1pt}
\end{equation}
where $\zeta = \sum\nolimits_{g = 1}^G { {{q_{g,{\tilde k}}}} }{{\bf{w}}_{\tilde k}^{{\rm{FD,H}}}{\bf{R}}_{g,\tilde k}^{\rm{J}}} {\bf{w}}_{\tilde k}^{{\rm{FD}}} +  {\rm{E}}{{\rm{N}}^{\rm UB}_{\tilde k}} +{\rm{Q}}{{\rm{E}}^{\rm UB}_{\tilde k}}+ \sigma _{\tilde k}^2$ is a constant independent of ${{\bf{f}}_k^{{\rm{FD}}}}$.

The constraint of power is like $L_2$-norm form, thus the project function is expressed as
\begin{equation}\label{20}
\setlength\abovedisplayskip{1pt}
{\mathcal{P}}\left({\bf{f}}_{l,k}^{{\rm{FD}}}\right)=\left\{ {\begin{array}{*{20}{c}}
{{\bf{f}}_{l,k}^{{\rm{FD}}},}&{\sum\nolimits_{k = 1}^K {\left\| {{\bf{f}}_{l,k}^{{\rm{FD}}}} \right\|_2^2}  \le {P_{\max }}}\\
{\sqrt {\frac{{{P_{\max }}}}{{\sum\nolimits_{k = 1}^K {\left\| {{\bf{f}}_{l,k}^{{\rm{FD}}}} \right\|_2^2} }}} {\bf{f}}_{l,k}^{{\rm{FD}}},}&{otherwise},
\end{array}} \right.
\setlength\belowdisplayskip{1pt}
\end{equation}
and for each iteration, we have
\setlength\abovedisplayskip{1pt}
\begin{equation}\label{21}
{\bf{f}}_k^{{\rm{FD}}} = {\mathcal {P} }\left ({\bf{f}}_k^{{\rm{FD}}} + \lambda {\nabla _{{\bf{f}}_k^{{\rm{FD}}}}} \eta \right ),\forall k \in K,
\setlength\belowdisplayskip{1pt}
\end{equation}
where $\lambda$ is step size. Considering the matrix multiplication in \eqref{19}, the complexity of this low-complexity method is $O(L^2M^2+KLMM_{\rm U}+GM^2_{\rm U})$ per iteration.

\subsection{Hybrid Beamforming Design}
After deriving the full-digital beamforming, the factorization approach for designing the hybrid beamforming can be formulated as
\begin{equation}\label{22}
\setlength\abovedisplayskip{1pt}
\begin{aligned}
    &\mathop {\min }\limits_{{{\bf{F}}^{\rm{RF}}},\left\{ {{\bf{f}}_{k}^{\rm{BB}}} \right\}}  \sum\nolimits_{k = 1}^K{\left\| {{\bf{f}}_{k}^{\rm{FD}} - {{\bf{F}}^{\rm{RF}}}{{\bf{f}}_{k}^{\rm{BB}}}} \right\|_2^2} \\
    &{\rm{s.t.}}\:\: \eqref{8e},
\end{aligned}
\setlength\belowdisplayskip{1pt}
\end{equation}
and $\forall k \in K$
\begin{equation}\label{23}
\setlength\abovedisplayskip{1pt}
    \begin{aligned}
    &\mathop {\min }\limits_{{{\bf{W}}^{\rm{RF}}_{k}},\left\{ {{\bf{w}}_{k}^{\rm{BB}}} \right\}} {\left\| {{\bf{w}}_{k}^{\rm{FD}} - {{\bf{W}}^{\rm{RF}}_{k}}{{\bf{w}}_{k}^{\rm{BB}}}} \right\|_2^2}  \\
    &{\rm{s.t.}}\:\: \eqref{8e}.
\end{aligned}
\setlength\belowdisplayskip{1pt}
\end{equation}
According to the scheme in\cite{8765770}, analog beamforming can be addressed and by solving \eqref{22} and \eqref{23}, and digital beamforming can be obtained. {After obtaining the analog beamformings of each AP, CPU uses the quantizer to process them and sends them to each AP.} The complexity of hybrid beamforming is $O(L^2M^2M_{\rm U}+KLMM^2_{\rm U}+L^2M^2N_{\rm RF}+KM^2_{\rm U}M_{\rm RF})$. More details are referred to \cite[Sec. 3]{8765770}.

\subsection{Convergence Analysis}
We introduce a superscript $t$ to each variable as the alternation index. In the $t$–th alternation, we can obtain the SINR value of each UE as ${\rm{SIN}}{{\rm{R}}_k}\left( {{\bf{w}}_k^{{\rm{BB,}}t},{\bf{W}}_k^{{\rm{RF,}}t},{\bf{f}}_{l,k}^{{\rm{BB,}}t},{\bf{F}}_l^{{\rm{RF,}}t}|q_{g,k}^t} \right)$, where ${\bf{w}}_k^{{\rm{BB,}}t}$, ${\bf{W}}_k^{{\rm{RF,}}t}$ are calculated by \eqref{13}, \eqref{23}, and ${\bf{f}}_{l,k}^{{\rm{BB,}}t},{\bf{F}}_l^{{\rm{RF,}}t}$ are calculated by \eqref{21}, \eqref{22}. ${q_{g,k}^t}$ is calculated by binary search. Then, we have
\begin{equation}
\begin{aligned}
&{\rm{SIN}}{{\rm{R}}_k}\left( {{\bf{w}}_k^{{\rm{BB,}}t},{\bf{W}}_k^{{\rm{RF,}}t},{\bf{f}}_{l,k}^{{\rm{BB,}}t},{\bf{F}}_l^{{\rm{RF,}}t}|q_{g,k}^t} \right)\\
 \le &{\rm{SIN}}{{\rm{R}}_k}\left( {{\bf{w}}_k^{{\rm{BB,}}t + 1},{\bf{W}}_k^{{\rm{RF,}}t + 1},{\bf{f}}_{l,k}^{{\rm{BB,}}t + 1},{\bf{F}}_l^{{\rm{RF,}}t + 1}|q_{g,k}^t} \right),
\end{aligned}
\end{equation}
The ${{\rm SINR}_k}$ increases as the beamforming is determined. Given the threshold of SINR, the maximum resistible jamming power can be solved by the binary search. As the value of SINR increases, the value of ${q_{g,k}^t}$ obtained from the binary search also increases. Thus, we can get ${q_{g,k}^{t + 1} \ge q_{g,k}^t}$ and for each alternation and the value of ${q_{g,k}}$ would increase or not decrease at least. Thus, the maximum resistible jamming power in \eqref{8} is monotonically non-decreasing after each alternation, and finally converge to a stationary point.

\subsection{The Overall Flow of Proposed Scheme}
After determining receiving beamforming and transmitting beamforming, the resistible jamming power can be solved by binary search. Assuming that the maximum number of alternations is $T$ and the convergence threshold is $\kappa$, the variables are alternately optimized using the above methodology. Algorithm 1 shows the general flow of the scheme. For each alternation, the total complexity of the AO-AJHBF scheme is $O(I(L^2M^2+KLMM_{\rm U}+GM^2_{\rm U})+KM^3_{\rm U}+KL^2M^2M_{\rm U}+KLMM^2_{\rm U}+L^2M^2N_{\rm RF}+KM^2_{\rm U}M_{\rm RF})$, where $I$ is the number of inner iterations.
\begin{algorithm}[h!]\label{alg_1}
\caption{Alternate Optimization Based Anti-jamming Hybrid Beamforming Design (AO-AJHBF).}\label{alg:alg1}
\begin{algorithmic}[1]
\STATE{\textbf{Input}: ${{{\bf{\bar H}}}_{l,k}}$,$~{{\bf Q}_{k}}$,$~{{{\bf{R}}}_{g,k}},~{\omega _k},~ \forall l \in L,~ \forall k \in K,~ \forall g \in G$}
\STATE{\textbf{Initialize}: ${{{\bf{F}}}^{\rm RF}_{l}}$,$~{{{\bf{f}}}^{\rm BB}_{l,k}}$,$~{{{\bf{W}}}^{\rm RF}_{k}}$,$~{{{\bf{w}}}^{\rm BB}_{k}}$,$~q_{g,k}$}
\STATE{${\mathbf{while}}$ $(t \le T~\mathbf{or}~q^{t+1}_{g,k}-q^{t}_{g,k}\le \kappa)$}
\STATE \hspace{0.5cm}{Calculate ${{{\bf{W}}}^{{\rm RF},t+1}_{k}}$,$~{{{\bf{w}}}^{{\rm BB},t+1}_{k}}$ by \eqref{13}, \eqref{23}, given ${{{\bf{F}}}^{{\rm RF},t}_{l}}$,$~{{{\bf{f}}}^{{\rm BB},t}_{l,k}}$,$~q^{t}_{g,k}$}
\STATE \hspace{0.5cm}{Calculate ${\bf{f}}_k^{{\rm{FD}},t+1}$ by \eqref{21} iteratively until convergence of $\eta  \left( {\bf{f}}_k^{{\rm{FD}},t+1} \right)$ then calculate ${{{\bf{F}}}^{{\rm RF},t+1}_{l}}$,$~{{{\bf{f}}}^{{\rm BB},t+1}_{l,k}}$ by \eqref{22}, given ${{{\bf{W}}}^{{\rm RF},t+1}_{k}}$,$~{{{\bf{w}}}^{{\rm BB},t+1}_{k}}$,$~q^{t}_{g,k}$}
\STATE \hspace{0.5cm}{Calculate $q^{t+1}_{g,k}$ by binary search, given ${{{\bf{W}}}^{{\rm RF},t+1}_{k}}$,$~{{{\bf{w}}}^{{\rm BB},t+1}_{k}}$,$~{{{\bf{F}}}^{{\rm RF},t+1}_{l}}$,$~{{{\bf{f}}}^{{\rm BB},t+1}_{l,k}}$}
\STATE {\textbf{Return}: ${{{\bf{F}}}^{\rm RF}_{l}}$,$~{{{\bf{f}}}^{\rm BB}_{l,k}}$,$~{{{\bf{W}}}^{\rm RF}_{k}}$,$~{{{\bf{w}}}^{\rm BB}_{k}}$}
\end{algorithmic}
\label{alg1}
\end{algorithm}


\section{Simulation Results}
In this section, we evaluate the performance of the proposed scheme. We assume that $K=5$ UEs and $G=2$ jammers are uniformly distributed in a cubic region with a side length of 1km. $L=3$ APs are located at one side of the region uniformly. The UPA of AP is located in the ${yoz}$ plane, while the UPA of UE is parallel to the ${yoz}$ plane towards AP. And we set $M=M_{\rm J}=36$, $N_{\rm RF}=18$, $M_{\rm U}=16$, $M_{\rm RF}=8$. For the channel model, we set the number of paths $P=3$, the small scale fading ${\alpha^p} \sim {\mathcal {CN}}\left( {0,1} \right)$, angle spread $\Delta \sim {\mathcal {U}}\left[ {-5^\circ,5^\circ} \right]$ and $\sigma^2=-107dBm$. The large scale fading $\beta$ is similar to\cite{elhoushy2021cell}. The other parameters are set as follows: $d=\lambda /2$, $\delta=-4$ and ${\gamma _{th}}=0dB$. {We choose the 4-bit quantizer thus ${\alpha}=0.990503$.} Considering the channel estimation error, the MMSE estimation approach is selected, with a NMSE of 0.01.

\begin{figure*}[htbp]
\centering
\subfloat[]{\includegraphics[scale=0.27]{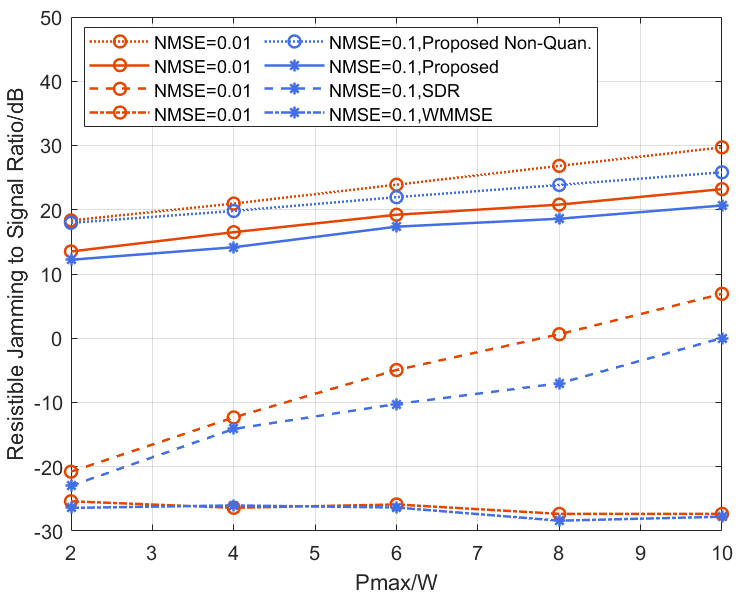}}
\subfloat[]{\includegraphics[scale=0.27]{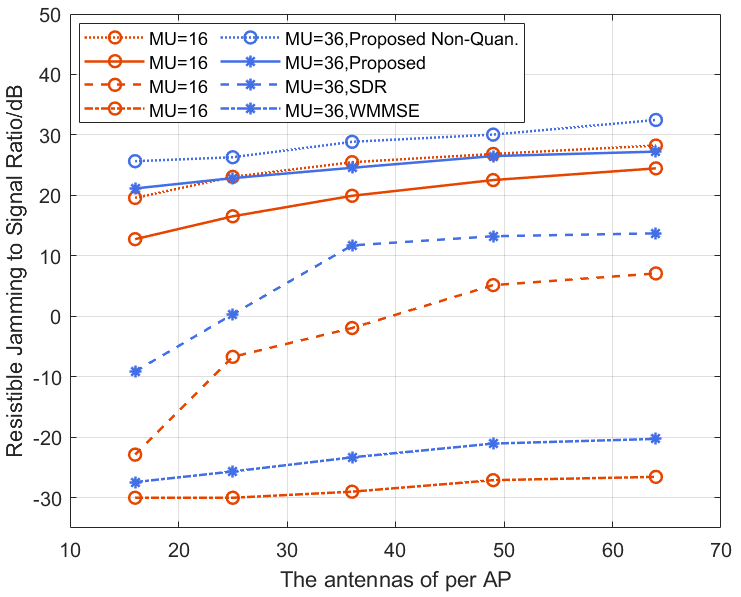}}
\subfloat[]{\includegraphics[scale=0.27]{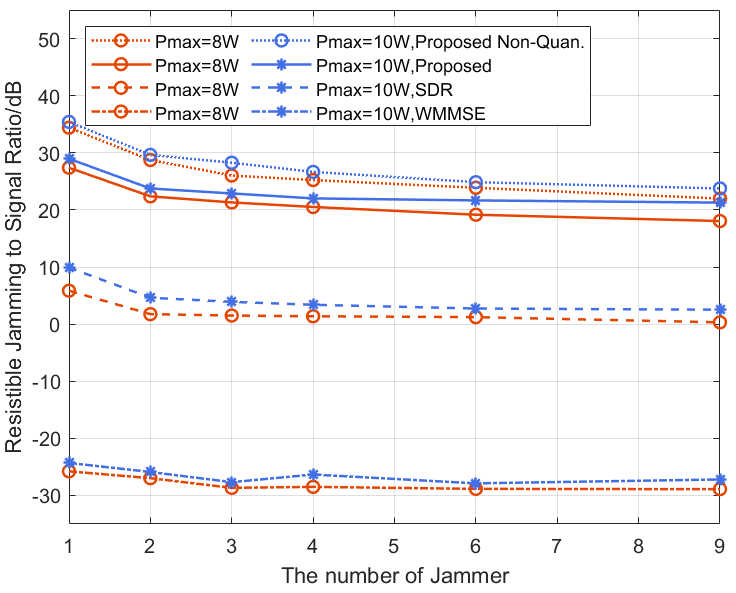}}
\setlength\abovedisplayskip{1pt}
\caption{(a) The average resistible JSR of different channel estimation error. (b) The average resistible JSR with respect to the number of antenna per AP, ${P_{\rm max}=8W}$. (c) The average resistible JSR with respect to the number of distributed jammers.}
\label{fig_2}
\end{figure*}

{We choose WMMSE scheme and SDR-based scheme as benchmarks (more details can be referred to Appendix B). The WMMSE is modified by the incorporation of jamming terms. The performance of proposed scheme without quantization is also compared.} The number of the maximum alternations for all algorithms is set to 3. The experimental apparatus employed a 64-bit Intel(R) Core(TM) i5-8400 CPU @2.80GHz with 8GB of RAM. Table \ref{tab1} presents the elapsed time for each scheme during a single run. Fig.1 (a) displays the average resistible jamming-to-signal ratio (JSR) for various channel estimation error. The proposed low-complexity algorithm outperforms the two benchmarks in terms of the average resistible JSR. It can be observed that an increase in channel estimation error results in a decline in the anti-jamming performance of the system. Fig.1 (b) depicts the average resistible JSR as a function of the number of antennas per AP with fixed ${M_{\rm U}}$. The results demonstrate that the anti-jamming effect of the system improves as the number of antennas per AP increases. The increase in the number of AP antennas allows for greater array gain at the transmitter side and more efficient inter-user interference elimination. Furthermore, the increase in the number of UE antennas allows for more effective spatial filtering, leading to a greater anti-jamming performance. 
\setlength\abovedisplayskip{1pt}
\begin{table}[htbp]
\caption{Single runtime of each scheme}
\begin{center}
\begin{tabular}{|c|c|c|c|}
\hline
\textbf{Scheme} & \textbf{{Proposed}}& \textbf{{SDR-based}}& \textbf{{WMMSE}} \\
\hline
\textbf{Runtime}& 0.48s&40.39s &0.42s  \\
\hline
\end{tabular}
\label{tab1}
\end{center}
\end{table}
\setlength\belowdisplayskip{1pt}

Subsequently, the impact of degree of spatial distribution of jammers on the anti-jamming performance of CF systems is investigated. It is assumed that the total number of antennas of all the randomly distributed jammers in the region remains constant, i.e., 36, and that all antennas are deployed equally to all jammers. {Fig.1 (c) illustrates that as the degree of spatial distribution increases, the average resistable JSR gradually decreases but our scheme demonstrates superior performance. This indicates that distributed jammers exert a more pronounced influence on the CF system compared to a single jammer, even when the total number of antennas is identical. The agent has limited capacity to prevent jamming from more different directions. Therefore, distributed jammers are more detrimental to communication systems.}

\section{Conclusion}
This letter proposed an anti-jamming transmission scheme AO-AJHBF for downlink mmWave CF-MU-MIMO systems. Exploiting the second-order statistics of jamming channel, the generalized Rayleigh quotient was used to design the receiving beamforming for resisting jamming. The transmitting beamforming was designed according to the max-min fairness principle, and we proposed a low-complexity method based on PGA to address the high complexity of the SDR-based method. Simulation results showed that the proposed scheme achieved superior anti-jamming performance.

\section{APPENDIX A: More details of Channel Estimation Error And Channel Quantization Error}
\subsection{Channel Estimation Error}
In this letter, we adopt the Saleh-Valenzuela channel model\cite{10013728}, which can be expressed as
\[{{\bf{H}}_{l,k}} = \sum\limits_{p = 1}^P {\sqrt {\beta _{l,k}^p} \alpha _{l,k}^p\left( {{{\bf{a}}_r}} \right)_{l,k}^p{{\left( {\left( {{{\bf{a}}_t}} \right)_{l,k}^p} \right)}^{\rm{H}}}}  = {\bf{U}}_{l,k}^r\left( {{{\bf{G}}_{l,k}} \odot {{\bf{S}}_{l,k}}} \right){\bf{U}}_{l,k}^{t,{\rm{H}}},\]
where ${\bf{U}}_{l,k}^r = \left[ {\left( {{{\bf{a}}_r}} \right)_{l,k}^1,\left( {{{\bf{a}}_r}} \right)_{l,k}^2, \cdots ,\left( {{{\bf{a}}_r}} \right)_{l,k}^P} \right]$, ${\bf{U}}_{l,k}^t = \left[ {\left( {{{\bf{a}}_t}} \right)_{l,k}^1,\left( {{{\bf{a}}_t}} \right)_{l,k}^2, \cdots ,\left( {{{\bf{a}}_t}} \right)_{l,k}^P} \right]$, ${{\bf{G}}_{l,k}} = diag$ $\left( {\sqrt {\beta _{l,k}^1} ,\sqrt {\beta _{l,k}^2} , \cdots ,\sqrt {\beta _{l,k}^P} } \right)$ and ${{\bf{S}}_{l,k}} = diag\left( {\alpha _{l,k}^1,\alpha _{l,k}^2, \cdots ,\alpha _{l,k}^P} \right)$. The small scale fading is supposed to be i.i.d. $CN\left( {0,1} \right)$. Then, we can obtain the auto-relation matrix of ${{\bf{H}}_{l,k}}$ , which is
\begin{equation*}
\begin{aligned}
{{\bf{R}}_{l,k}} &= {\rm{E}}\left\{ {vec\left( {{{\bf{H}}_{l,k}}} \right)ve{c^{\rm{H}}}\left( {{{\bf{H}}_{l,k}}} \right)} \right\}\\
&= {\rm{E}}\left\{ {\left( {{\bf{U}}_{l,k}^{t,*} \otimes {\bf{U}}_{l,k}^r} \right)vec\left( {{{\bf{G}}_{l,k}} \odot {{\bf{S}}_{l,k}}} \right)ve{c^{\rm{H}}}\left( {{{\bf{G}}_{l,k}} \odot {{\bf{S}}_{l,k}}} \right){{\left( {{\bf{U}}_{l,k}^{t,*} \otimes {\bf{U}}_{l,k}^r} \right)}^{\rm{H}}}} \right\}\\
&= {\rm{E}}\left\{ {\left( {{\bf{U}}_{l,k}^{t,*} \otimes {\bf{U}}_{l,k}^r} \right)diag\left[ {vec\left( {{{\bf{G}}_{l,k}} \odot {{\bf{G}}_{l,k}}} \right)} \right]{{\left( {{\bf{U}}_{l,k}^{t,*} \otimes {\bf{U}}_{l,k}^r} \right)}^{\rm{H}}}} \right\}. 
\end{aligned}
\end{equation*}
And we can also get that
\[\begin{array}{l}
{\rm{E}}\left\{ {vec\left( {{{\bf{H}}_{l,k}}} \right)ve{c^{\rm{H}}}\left( {{{\bf{H}}_{j,k}}} \right)} \right\}\\
 = {\rm{E}}\left\{ {\left( {{\bf{U}}_{l,k}^{t,*} \otimes {\bf{U}}_{l,k}^r} \right)vec\left( {{{\bf{G}}_{l,k}} \odot {{\bf{S}}_{l,k}}} \right)ve{c^{\rm{H}}}\left( {{{\bf{G}}_{j,k}} \odot {{\bf{S}}_{j,k}}} \right){{\left( {{\bf{U}}_{j,k}^{t,*} \otimes {\bf{U}}_{j,k}^r} \right)}^{\rm{H}}}} \right\}\\
 = 0, (l \neq j).
\end{array}\]
Thus, the auto-relation matrix of ${{\bf{H}}_k}$ can be expressed as
\[{{\bf{R}}_k} = {\rm{E}}\left\{ {vec\left( {{{\bf{H}}_k}} \right)ve{c^{\rm{H}}}\left( {{{\bf{H}}_k}} \right)} \right\} = blkdiag\left( {{{\bf{R}}_{1,k}},{{\bf{R}}_{2,k}}, \cdots ,{{\bf{R}}_{L,k}}} \right).\]
By sending the pilot matrix to AP, we can get the receiving signal at AP-$l$ , which is
\[{{\bf{Y}}_l} = \sum\limits_{k = 1}^K {{{\bf{H}}_{l,k}}{{\bf{F}}_k}{\bf{\Phi }}_k^{\rm{T}}}  + {{\bf{N}}_l},\]
where ${{\bf{F}}_k}$ is the precoding of UE-$k$ when sending the pilot. And $\Phi _k$ is the pilot matrix of UE-$k$ which satisfies $\Phi _k^{\rm{H}}{\Phi _m} = {\tau _p}{\bf{I}}$ when $k=m$ and ${\bf{0}}$ otherwise. ${{\bf{N}}_l} \sim CN\left( {0,{\sigma ^2}{\bf{I}}} \right)$.
We can derive ${{\bf{H}}_{l,k}}$ by the projection of ${{\bf{Y}}_l}$ onto ${\bf{\Phi }}_k^*$, which is
\[\begin{array}{c}
{{\bf{Y}}_{l,k}} = {{\bf{Y}}_l}{\bf{\Phi }}_k^* = \sum\limits_{k = 1}^K {{{\bf{H}}_{l,k}}{{\bf{F}}_k}{\bf{\Phi }}_k^{\rm{T}}{\bf{\Phi }}_k^*}  + {{\bf{N}}_l}{\bf{\Phi }}_k^*\\
 = {\tau _p}{{\bf{H}}_{l,k}}{{\bf{F}}_k} + {{\bf{N}}_l}{\bf{\Phi }}_k^*.
\end{array}\]
Then, by using MMSE estimation, we can get the estimated channel as
\[{{\bf{\hat h}}_{l,k}} = vec\left( {{{{\bf{\hat H}}}_{l,k}}} \right) = {{\bf{R}}_{l,k}}{\bf{\tilde F}}_k^{\rm{H}}{\bf{\Psi }}_{l,k}^{ - 1}{{\bf{y}}_{l,k}},\]
where ${\bf{\tilde F}}_k^{} = {\bf{F}}_k^{\rm{T}} \otimes {\bf{I}}$, ${\bf{\Psi }}_{l,k}^{} = {\tau _p}{\bf{\tilde F}}_k^{}{{\bf{R}}_{l,k}}{\bf{\tilde F}}_k^{\rm{H}} + {\sigma ^2}{\bf{I}}$, ${{\bf{y}}_{l,k}} = vec\left( {{{\bf{Y}}_{l,k}}} \right) = {\tau _p}{\bf{\tilde F}}_k^{}{{\bf{h}}_{l,k}} + {{\bf{n}}_l}$ and ${{\bf{n}}_l} = vec\left( {{{\bf{N}}_l}{\bf{\Phi }}_k^*} \right)$. Thus, the estimation error ${{\bf{\tilde h}}_{l,k}} = {{\bf{h}}_{l,k}} - {{\bf{\hat h}}_{l,k}}$ and ${{\bf{\hat h}}_{l,k}}$ are independent. And we can get the auto-relation matrix of ${{\bf{\hat h}}_{l,k}}$, which is ${{\bf{\hat R}}_{l,k}} \buildrel \Delta \over = {\tau _p}{{\bf{R}}_{l,k}}{\bf{\tilde F}}_k^{\rm{H}}{\bf{\Psi }}_{l,k}^{ - 1}{\bf{\tilde F}}_k^{}{{\bf{R}}_{l,k}}$. Then, the auto-relation matrix of ${{\bf{\hat h}}_k}$ is ${{\bf{\hat R}}_k} = blkdiag\left( {{{{\bf{\hat R}}}_{1,k}},{{{\bf{\hat R}}}_{2,k}}, \cdots ,{{{\bf{\hat R}}}_{L,k}}} \right)$. Finally, the auto-relation matrix of estimation error ${{\bf{\tilde h}}_k}$ is ${{\bf{Q}}_k} = {{\bf{R}}_k} - {{\bf{\hat R}}_k}$. And in practice, ${{\bf{R}}_k}$ could be obtained by statistically averaging the historical data of CSI thus ${{\bf{Q}}_k}$ can be calculated and it is possible to obtain the ${\lambda _{\max }}\left( {{{\bf{Q}}_k}} \right)$.

The entire procedure of Eq. (10) is as below:
\begin{equation*}
    \begin{aligned}
{\rm{E}}{{\rm{N}}_k} &= \sum\limits_{j = 1}^K {\rm{E}} \left\{ {{{\left| {{\bf{w}}_k^{{\rm{FD,H}}}{{{\bf{\tilde H}}}_k}{\bf{f}}_j^{{\rm{FD}}}} \right|}^2}} \right\} = \sum\limits_{j = 1}^K {\rm{E}} \left\{ {{\bf{w}}_k^{{\rm{FD,H}}}{{{\bf{\tilde H}}}_k}{\bf{f}}_j^{{\rm{FD}}}{\bf{f}}_j^{{\rm{FD,H}}}{\bf{\tilde H}}_k^{{\rm{H}}}{\bf{w}}_k^{{\rm{FD}}}} \right\}\\
&= \sum\limits_{j = 1}^K {\rm{E}} \left\{ {\left( {{\bf{f}}_j^{{\rm{FD,T}}} \otimes {\bf{w}}_k^{{\rm{FD,H}}}} \right)vec\left( {{{{\bf{\tilde H}}}_k}} \right)ve{c^{\rm{H}}}\left( {{{{\bf{\tilde H}}}_k}} \right){{\left( {{\bf{f}}_j^{{\rm{FD,T}}} \otimes {\bf{w}}_k^{{\rm{FD,H}}}} \right)}^{\rm{H}}}} \right\}\\
&= \sum\limits_{j = 1}^K {\left( {{\bf{f}}_j^{{\rm{FD,T}}} \otimes {\bf{w}}_k^{{\rm{FD,H}}}} \right){\rm{E}}\left\{ {vec\left( {{{{\bf{\tilde H}}}_k}} \right)ve{c^{\rm{H}}}\left( {{{{\bf{\tilde H}}}_k}} \right)} \right\}{{\left( {{\bf{f}}_j^{{\rm{FD,T}}} \otimes {\bf{w}}_k^{{\rm{FD,H}}}} \right)}^{\rm{H}}}} \\
&= \sum\limits_{j = 1}^K {\left( {{\bf{f}}_j^{{\rm{FD,T}}} \otimes {\bf{w}}_k^{{\rm{FD,H}}}} \right){{\bf{Q}}_k}{{\left( {{\bf{f}}_j^{{\rm{FD,T}}} \otimes {\bf{w}}_k^{{\rm{FD,H}}}} \right)}^{\rm{H}}}} ,
    \end{aligned}
\end{equation*}
where ${{\bf{Q}}_k} = {\rm{E}}\left\{ {vec\left( {{{{\bf{\tilde H}}}_k}} \right)ve{c^{\rm{H}}}\left( {{{{\bf{\tilde H}}}_k}} \right)} \right\} \in {C^{M{M_U} \times M{M_U}}}$. Let ${{\bf{s}}_j} = {\bf{f}}_j^{{\rm{FD,T}}} \otimes {\bf{w}}_k^{{\rm{FD,H}}} \in {C^{1 \times M{M_U}}}$, the ${\rm{E}}{{\rm{N}}_k}$ can be reformulated as ${\rm{E}}{{\rm{N}}_k} = \sum\limits_{j = 1}^K {{{\bf{s}}_j}{{\bf{Q}}_k}{\bf{s}}_j^{\rm{H}}}$. Noting that by the nature of norm, we can obtain ${\left\| {{{\bf{s}}_j}} \right\|_2} = {\left\| {{\bf{f}}_j^{{\rm{FD,T}}} \otimes {\bf{w}}_k^{{\rm{FD,H}}}} \right\|_2} = {\left\| {{\bf{f}}_j^{{\rm{FD,T}}}} \right\|_2}{\left\| {{\bf{w}}_k^{{\rm{FD,H}}}} \right\|_2} = {\left\| {{\bf{f}}_j^{{\rm{FD}}}} \right\|_2}$. Therefore, the ${\rm{E}}{{\rm{N}}_k}$ is converted as ${\rm{E}}{{\rm{N}}_k} = \sum\limits_{j = 1}^K {{{\bf{s}}_j}{{\bf{Q}}_k}{\bf{s}}_j^{\rm{H}}}  = \sum\limits_{j = 1}^K {\left\| {{\bf{f}}_j^{{\rm{FD}}}} \right\|_2^2{{\bf{\beta }}_j}{{\bf{Q}}_k}{\bf{\beta }}_j^{\rm{H}}}$, where ${{\bf{\beta }}_j} = \frac{{{{\bf{s}}_j}}}{{{{\left\| {{\bf{f}}_j^{{\rm{FD}}}} \right\|}_2}}}$. 

Let ${\bf{p}} = {\left[ {\left\| {{\bf{f}}_1^{{\rm{FD}}}} \right\|_2^2,\left\| {{\bf{f}}_2^{{\rm{FD}}}} \right\|_2^2, \cdots ,\left\| {{\bf{f}}_K^{{\rm{FD}}}} \right\|_2^2} \right]^{\rm{T}}}$ and ${{\bf{q}}_k} = {\left[ {{{\bf{\beta }}_1}{{\bf{Q}}_k}{\bf{\beta }}_1^{\rm{H}},{{\bf{\beta }}_2}{{\bf{Q}}_k}{\bf{\beta }}_2^{\rm{H}}, \cdots ,{{\bf{\beta }}_K}{{\bf{Q}}_k}{\bf{\beta }}_K^{\rm{H}}} \right]^{\rm{T}}}$, then ${\rm{E}}{{\rm{N}}_k}$ is conducted as below:
\begin{equation*}
    \begin{aligned}
{\rm{E}}{{\rm{N}}_k} &= {{\bf{p}}^{\rm{T}}}{{\bf{q}}_k} = {\left\| {{{\bf{p}}^{\rm{T}}}{{\bf{q}}_k}} \right\|_1}\\
 &\le {\left\| {\bf{p}} \right\|_1}{\left\| {{{\bf{q}}_k}} \right\|_1} = L{P_{\max }}\sum\limits_{j = 1}^K {{{\bf{\beta }}_j}{{\bf{Q}}_k}{\bf{\beta }}_j^{\rm{H}}} \\
 &\le LK{P_{\max }} \cdot {\lambda _{\max }}\left( {{{\bf{Q}}_k}} \right) = {\rm{EN}}_k^{{\rm{UB}}}.
    \end{aligned}
\end{equation*}

\subsection{Channel Quantization Error}
Considering the limited capacity of fronthaul link in cell free system, we suppose that the channel is first quantized at each AP and then transmitted to CPU\cite{9658000}, \cite{8730536}. The quantization can be expressed as
\[{{\bf{\bar h}}_{l,k}} = {\mathrm Q}\left( {{{{\bf{\hat h}}}_{l,k}}} \right) = \alpha {{\bf{\hat h}}_{l,k}} + {{\bf{\bar q}}_{l,k}},\]
where $\alpha $ is a distortion factor corresponding to quantization bits per sample. 
And ${{\bf{\bar q}}_{l,k}}$ is the quantization error, which satisfies ${{\bf{\bar q}}_{l,k}} \sim CN\left( {0,\sigma _{q,l,k}^2{\bf{I}}} \right)$.
Then, we can derive that
\[\sigma _{q,l,k}^2 = \alpha \left( {1 - \alpha } \right)\frac{{{\tau _p}{\rho _p}\beta _{l,k}^2}}{{{\tau _p}{\rho _p}\sum\limits_{j \ne k}^K {\beta _{l,j}^{}{{\left| {\varphi _k^{\rm{H}}{\varphi _j}} \right|}^2}}  + 1}},\]
where ${\tau _p}$ is the training length, ${\rho _p}$ is the transmitted power, $\beta _{l,j}^{}$ is the large scale fading factor and ${\varphi _j}$ is the pilot vector. Then, the quantization error of ${{\bf{\hat H}}_k}$ is expressed as 
\[{{\bf{\Sigma }}_k}=[{{\bf{\Sigma }}_{1,k}}, {{\bf{\Sigma }}_{2,k}}, \dots, {{\bf{\Sigma }}_{L,k}}],\]
where ${{\bf{\Sigma }}_{l,k}} \sim CN\left( {0,\sigma _{q,l,k}^2{\bf{I}}} \right)$.

The relationship of $\alpha $ and the number of quantization bit is expressed as Table 1\cite{9658000}.
\begin{table}[h]
\caption{The relationship of $\alpha $ and the number of quantization bit}
\begin{center}
\begin{tabular}{|c|c|c|c|c|c|}
\hline
\textbf{Bits} & \textbf{{1}}& \textbf{{2}}& \textbf{{3}} &\textbf{{4}}& \textbf{{5}} \\
\hline
\textbf{$\alpha$}& 0.6366 &0.8825 &0.96546 &0.990503 &0.997501  \\
\hline
\end{tabular}
\label{tab1}
\end{center}
\end{table}

Note that the detail of more proof can be referred to\cite{9658000}, \cite{8730536}, therefore we only offer some significant procedures.

Then, similar to the way of treating channel estimation error, we can get the channel quantization error term, which is
\begin{equation*}
    \begin{aligned}
{\rm{Q}}{{\rm{E}}_k} &= \sum\limits_{j = 1}^K {\rm{E}} \left\{ {{{\left| {{\bf{w}}_k^{{\rm{FD,H}}}{{\bf{\Sigma }}_k}{\bf{f}}_j^{{\rm{FD}}}} \right|}^2}} \right\} = \sum\limits_{j = 1}^K {\rm{E}} \left\{ {{\bf{w}}_k^{{\rm{FD,H}}}{{\bf{\Sigma }}_k}{\bf{f}}_j^{{\rm{FD}}}{\bf{f}}_j^{{\rm{FD,H}}}{\bf{\Sigma }}_k^{\rm{H}}{\bf{w}}_k^{{\rm{FD}}}} \right\}\\
&= \sum\limits_{j = 1}^K {\rm{E}} \left\{ {\left( {{\bf{f}}_j^{{\rm{FD,T}}} \otimes {\bf{w}}_k^{{\rm{FD,H}}}} \right)vec\left( {{{\bf{\Sigma }}_k}} \right)ve{c^{\rm{H}}}\left( {{{\bf{\Sigma }}_k}} \right){{\left( {{\bf{f}}_j^{{\rm{FD,T}}} \otimes {\bf{w}}_k^{{\rm{FD,H}}}} \right)}^{\rm{H}}}} \right\}\\
&= \sum\limits_{j = 1}^K {\left( {{\bf{f}}_j^{{\rm{FD,T}}} \otimes {\bf{w}}_k^{{\rm{FD,H}}}} \right){\rm{E}}\left\{ {vec\left( {{{\bf{\Sigma }}_k}} \right)ve{c^{\rm{H}}}\left( {{{\bf{\Sigma }}_k}} \right)} \right\}{{\left( {{\bf{f}}_j^{{\rm{FD,T}}} \otimes {\bf{w}}_k^{{\rm{FD,H}}}} \right)}^{\rm{H}}}} \\
&\le L{P_{\max }}K{\omega _k} = {\rm{QE}}_k^{{\rm{UB}}},
    \end{aligned}
\end{equation*}
where ${\omega _k} = \max \left( {\left[ {\sigma _{q,1,k}^2,\sigma _{q,2,k}^2, \cdots ,\sigma _{q,L,k}^2} \right]} \right)$.

\section{APPENDIX B: More details of SDR-based and WMMSE algorithm}
\subsection{SDR-based scheme}
The SDR-based algorithm is mainly to address the (14). Through the SDR method, (14) can be transformed into a convex problem, which is expressed as (15). For more details, (15) can be further expressed as
\[\begin{array}{*{20}{l}}
&{\mathop {\max }\limits_{\left\{ {{\bf{U}}_k^{{\rm{FD}}}} \right\},\varepsilon } \varepsilon }\\
{{\rm{s}}.{\rm{t}}.\:\:}&{\sum\limits_{k = 1}^K {Tr\left\{ {{\bf{U}}_k^{{\rm{FD}}}{{\bf{S}}_l}} \right\}}  \le {P_{\max }},\forall l \in L,}\\
&{\bf{U}}_k^{{\rm{FD}}} \succeq 0,{\rm{rank}}\left( {{\bf{U}}_k^{{\rm{FD}}}} \right) = 1,\forall k \in K,\\
&\frac{{{\bf{w}}_k^{{\rm{FD,H}}}{{{\bf{\bar H}}}_k}{\bf{U}}_k^{{\rm{FD}}}{\bf{\bar H}}_k^{\rm{H}}{\bf{w}}_k^{\rm{H}}}}{{\sum\limits_{j \ne k}^K {{\bf{w}}_k^{{\rm{FD,H}}}{{{\bf{\bar H}}}_k}{\bf{U}}_j^{{\rm{FD}}}{\bf{\bar H}}_k^{\rm{H}}{\bf{w}}_k^{\rm{H}}}  + {\rm EN}_k^{\rm UB} +{\rm{QE}}_k^{{\rm{UB}}}+ \sum\limits_{g = 1}^G {{q_{g,k}}{\bf{w}}_k^{{\rm{FD,H}}}{R_{g,k}}{\bf{w}}_k^{\rm{H}}}  + \sigma _k^2}} \ge \varepsilon. 
\end{array}\]

Except for the constraint ${\rm{rank}}\left( {{\bf{U}}_k^{{\rm{FD}}}} \right) = 1,\forall k \in K$ ,the problem is convex. And by utilizing the CVX toolbox, the problem can be solved and $\left\{ {{\bf{U}}_k^{{\rm{FD}}}} \right\}$ can be obtained. By performing the eigenvalue decomposition, we select the eigenvector corresponding to the largest eigenvalue of $\left\{ {{\bf{U}}_k^{{\rm{FD}}}} \right\}$ as the optimal full-digital beamforming. By replacing the step 5 of Algorithm 1 with the aforementioned method, the SDR-based scheme is generated. The complexity of solving this problem is $O(K^{3}L^{6.5}M^{6.5}{\rm log({1/\epsilon})})$ (where ${\epsilon}$ is the accuracy of the solution)\cite{ye2011interior}.

\subsection{WMMSE-based scheme}
The WMMSE scheme is modified by the incorporation of jamming terms. In this text, we only provide the significant procedures that are related to the beamforming design. And the original WMMSE scheme can be referred to \cite{5756489} and \cite{feng2021weighted}. Firstly, the received signal of the $k$-th UE can be expressed as
\begin{equation*}
\begin{aligned}
{{\hat s}_k} &= {\bf{w}}_k^{\rm{H}}\sum\limits_{l = 1}^L {{{{\bf{\bar H}}}_{l,k}}{{\bf{f}}_{l,k}}{s_k}}  + {\bf{w}}_k^{\rm{H}}\sum\limits_{j \ne k}^K {\sum\limits_{l = 1}^L {{{{\bf{\bar H}}}_{l,k}}{{\bf{f}}_{l,j}}{s_j}} }  + {\bf{w}}_k^{\rm{H}}\sum\limits_{j = 1}^K {\sum\limits_{l = 1}^L {{{{\bf{\tilde H}}}_{l,k}}{{\bf{f}}_{l,j}}{s_j}} } \\
 &+ {\bf{w}}_k^{\rm{H}}\sum\limits_{j = 1}^K {\sum\limits_{l = 1}^L {{{{\bf{\Sigma }}_k}}{{\bf{f}}_{l,j}}{s_j}} }+ {\bf{w}}_k^{\rm{H}}\sum\limits_{g = 1}^G {\sqrt {{q_{g,k}}} {{\bf{J}}_{g,k}}{{\bf{w}}_{{\rm{J,}}g,k}}{z_g}}  + {\bf{w}}_k^{\rm{H}}{{\bf{n}}_k}.
\end{aligned}
\end{equation*}

The MSE for the $k$-th UE is calculated as
\begin{equation*}
\begin{aligned}
{E_k} &= {\rm{E}}\left\{ {\left( {{{\hat s}_k} - {s_k}} \right){{\left( {{{\hat s}_k} - {s_k}} \right)}^{\rm{H}}}} \right\} = \left( {{\bf{w}}_k^{\rm{H}}\sum\limits_{l = 1}^L {{{{\bf{\bar H}}}_{l,k}}{{\bf{f}}_{l,k}}}  - 1} \right){\left( {{\bf{w}}_k^{\rm{H}}\sum\limits_{l = 1}^L {{{{\bf{\bar H}}}_{l,k}}{{\bf{f}}_{l,k}}}  - 1} \right)^{\rm{H}}}\\
 &+ {\bf{w}}_k^{\rm{H}}\sum\limits_{j \ne k}^K {\left[ {\left( {\sum\limits_{l = 1}^L {{{{\bf{\bar H}}}_{l,k}}{{\bf{f}}_{l,j}}} } \right){{\left( {\sum\limits_{l = 1}^L {{{{\bf{\bar H}}}_{l,k}}{{\bf{f}}_{l,j}}} } \right)}^{\rm{H}}}} \right]} {{\bf{w}}_k} + {\bf{w}}_k^{\rm{H}}\left( {\sum\limits_{g = 1}^G {{q_{g,k}}{{\bf{R}}_{g,k}}} } \right){{\bf{w}}_k} + \hat \sigma _k^2,
\end{aligned}
\end{equation*}
where $\hat \sigma _k^2 = \sigma _k^2 + LK{P_{\max }} \cdot {\lambda _{\max }}\left( {{{\bf{Q}}_k}} \right) +LK{P_{\max }}{\omega _k}$.

According to the relevant procedure, we can get the ${\bf{w}_{k}}$, which can be expressed as
\begin{equation*}
\begin{aligned}
{{\bf{w}}_k} &= {\left\{ {\sum\limits_{j = 1}^K {\left[ {\left( {\sum\limits_{l = 1}^L {{{{\bf{\bar H}}}_{l,k}}{{\bf{f}}_{l,j}}} } \right){{\left( {\sum\limits_{l = 1}^L {{{{\bf{\bar H}}}_{l,k}}{{\bf{f}}_{l,j}}} } \right)}^{\rm{H}}}} \right]}  + \left( {\sum\limits_{g = 1}^G {{q_{g,k}}{{\bf{R}}_{g,k}}} } \right) + \hat \sigma _k^2{\bf{I}}} \right\}^{ - 1}}\left( {\sum\limits_{l = 1}^L {{{{\bf{\bar H}}}_{l,k}}{{\bf{f}}_{l,k}}} } \right)\\
&= {\bf{J}}_k^{ - 1}\left( {\sum\limits_{l = 1}^L {{{{\bf{\bar H}}}_{l,k}}{{\bf{f}}_{l,k}}} } \right),  
\end{aligned}
\end{equation*}
where ${{\bf{J}}_k} = \sum\limits_{j = 1}^K {\left[ {\left( {\sum\limits_{l = 1}^L {{{{\bf{\bar H}}}_{l,k}}{{\bf{f}}_{l,j}}} } \right){{\left( {\sum\limits_{l = 1}^L {{{{\bf{\bar H}}}_{l,k}}{{\bf{f}}_{l,j}}} } \right)}^{\rm{H}}}} \right]}  + \left( {\sum\limits_{g = 1}^G {{q_{g,k}}{{\bf{R}}_{g,k}}} } \right) + \hat \sigma _k^2{\bf{I}}$.

Then, by replacing the ${\bf{w}_{k}}$ in the MSE, we can obtain
\begin{equation*}
{E_k} = 1 - {\left( {\sum\limits_{l = 1}^L {{{{\bf{\bar H}}}_{l,k}}{{\bf{f}}_{l,k}}} } \right)^{\rm{H}}}{\bf{J}}_k^{ - 1}\left( {\sum\limits_{l = 1}^L {{{{\bf{\bar H}}}_{l,k}}{{\bf{f}}_{l,k}}} } \right) ,
\end{equation*}
and the optimal value $W_k^{*,opt}$ is $E_k^{ - 1}$.

Next, we continue solving the transmitting beamforming. Once we get the ${\bf{w}_{k}}$ and $W_k^{*,opt}$, the equivalent problem of maximization sum rate can be expressed as
\[\begin{array}{l}
\mathop {\min }\limits_{\left\{ {{{\bf{f}}_{l,k}}} \right\}} \sum\limits_{k = 1}^K {Tr\left( {{W_k}{E_k}} \right)} \\
s.t.\sum\limits_{k = 1}^K {Tr\left( {{\bf{f}}_{l,k}^{\rm{H}}{{\bf{f}}_{l,k}}} \right)}  \le {P_{\max }}.
\end{array}\]
Through the Lagrange Multiplier Method, we can obtain the Lagrange function, which can be written as
\begin{equation*}
\begin{aligned}
L\left( {{{\bf{f}}_{l,k}},{\lambda _l}} \right) &= \sum\limits_{k = 1}^K {Tr\left( {{W_k}{E_k}} \right)}  + \sum\limits_{l = 1}^L {{\lambda _l}\left[ {\sum\limits_{k = 1}^K {Tr\left( {{\bf{f}}_{l,k}^{\rm{H}}{{\bf{f}}_{l,k}}} \right)}  - {P_{\max }}} \right]} \\
 &= \sum\limits_{k = 1}^K {Tr\left[ {{W_k}{\bf{w}}_k^{\rm{H}}\left( {\sum\limits_{l = 1}^L {{{{\bf{\bar H}}}_{l,k}}{{\bf{f}}_{l,k}}} } \right){{\left( {\sum\limits_{l = 1}^L {{{{\bf{\bar H}}}_{l,k}}{{\bf{f}}_{l,k}}} } \right)}^{\rm{H}}}{{\bf{w}}_k}} \right]} \\
 &- \sum\limits_{k = 1}^K {Tr\left[ {{{\left( {\sum\limits_{l = 1}^L {{{{\bf{\bar H}}}_{l,k}}{{\bf{f}}_{l,k}}} } \right)}^{\rm{H}}}{{\bf{w}}_k}{W_k}} \right]}  - \sum\limits_{k = 1}^K {Tr\left[ {{W_k}{\bf{w}}_k^{\rm{H}}\left( {\sum\limits_{l = 1}^L {{{{\bf{\bar H}}}_{l,k}}{{\bf{f}}_{l,k}}} } \right)} \right]} \\
 &+ \sum\limits_{l = 1}^L {{\lambda _l}\left[ {\sum\limits_{k = 1}^K {Tr\left( {{\bf{f}}_{l,k}^{\rm{H}}{{\bf{f}}_{l,k}}} \right)}  - {P_{\max }}} \right]} ,
\end{aligned}
\end{equation*}
where is ${\lambda _l}$ the Lagrange multiplier of $l$-th constraint.

Then the gradient of $L\left( {{{\bf{f}}_{l,k}},{\lambda _l}} \right)$ with respect to ${{\bf{f}}_{l,k}}$ can be obtained, which is
\[{\nabla _{{{\bf{f}}_{l,k}}}}L = {\bf{\bar H}}_{l,k}^{\rm{H}}{{\bf{w}}_k}{W_k}{\bf{w}}_k^{\rm{H}}\left( {\sum\limits_{l = 1}^L {{{{\bf{\bar H}}}_{l,k}}{{\bf{f}}_{l,k}}} } \right) - {\bf{\bar H}}_{l,k}^{\rm{H}}{{\bf{w}}_k}{W_k} + {\lambda _l}{{\bf{f}}_{l,k}}.\]
By setting ${\nabla _{{{\bf{f}}_{l,k}}}}L = 0$, we can get the transforming beamforming, which can be represented as
\begin{equation*}
{{\bf{f}}_{l,k}} = {\left( {{\bf{\bar H}}_{l,k}^{\rm{H}}{{\bf{w}}_k}{W_k}{\bf{w}}_k^{\rm{H}}{{{\bf{\bar H}}}_{l,k}} + {\lambda _l}{\bf{I}}} \right)^{ - 1}}\left[ {{\bf{\bar H}}_{l,k}^{\rm{H}}{{\bf{w}}_k}{W_k} - {\bf{\bar H}}_{l,k}^{\rm{H}}{{\bf{w}}_k}{W_k}{\bf{w}}_k^{\rm{H}}{{{\bf{\bar H}}}_{l,k}}\left( {\sum\limits_{i \ne l}^L {{{{\bf{\bar H}}}_{i,k}}{{\bf{f}}_{i,k}}} } \right)} \right].    
\end{equation*}
The value of ${\lambda _l}$ can be determined by binary search.

Finally, by replacing the beamforming steps of Algorithm 1 with the WMMSE method, the WMMSE-based scheme is generated. The complexity is $O(K^2L^3M^3+K^2L^2M^2M_{\rm U}+K^2LMM^2_{\rm U}+KM^3_{\rm U})$ per iteration\cite{5756489}.

\bibliographystyle{IEEEtran}      
\bibliography{ref.bib}                        
\end{document}